\documentclass[sigconf, 10pt, nonacm]{acmart}
\usepackage{graphicx} 
\usepackage{subfig}
\usepackage{array}
\usepackage{xspace}
\usepackage{multirow}
\usepackage[normalem]{ulem}
\usepackage{soul}

\newif\ifcomments
\commentstrue
\ifcomments
\usepackage[textsize=small,color=lightgray]{todonotes}
\else
\usepackage[disable]{todonotes}
\fi

\newcommand{\sys}{\texttt{\mbox{MultiWorld}}\xspace}
\newcommand{\fref}[1]{Figure~\ref{#1}}

\newcommand{\secref}[1]{\S\ref{#1}}

\newcommand{\mypara}[1]{\noindent\textbf{#1}}

\newenvironment{tightitemize}%
 {\begin{list}{$\bullet$}{%
        \setlength{\leftmargin}{10pt}
        \setlength{\itemsep}{0pt}%
        \setlength{\parsep}{0pt}%
        \setlength{\topsep}{0pt}%
        \setlength{\parskip}{0pt}%
        }%
 }%
{\end{list}}

\usepackage{xcolor}

\definecolor{blue}{rgb}{0,0,1}
\definecolor{red}{rgb}{1,0,0}

\title[]{Enabling Elastic Model Serving with MultiWorld}
\author{Myungjin Lee, Akshay Jajoo, Ramana Rao Kompella}
\affiliation{%
  \institution{Cisco Research}%
  \country{}%
}

\renewcommand{\shortauthors}{}

\begin{abstract}
Machine learning models have been exponentially growing in terms of their parameter size over the past few years.
We are now seeing the rise of trillion-parameter models.
The large models cannot fit into a single GPU and thus require partitioned deployment across GPUs and even hosts.
A high-performance collective communication library (CCL) such as NCCL is essential to fully utilize expensive GPU resources.
However, CCL is not a great fit for inference.
Unlike training for which a fixed amount of GPU resources is used for fixed workloads (e.g., input datasets),
the inference workloads
can change dynamically over time. 
Failures at the serving time can also impact individual user's experiences directly.
In contrast, workers in a CCL process group share a single fault domain and the process group cannot grow as the workloads increase.
The gap between the unique characteristics of model serving and CCL's nature makes it hard to serve large models elastically.
To bridge the gap, we propose \sys that enables fault tolerance and online scaling at the granularity of workers for model serving.
Our evaluation showcases that enabling these new functionalities incurs small overheads (1.4-4.3\% throughput loss) for most of the scenarios we tested.
 
\end{abstract}

\begin{document}

\maketitle

\section{Introduction}
\label{s:intro}
The recent remarkable advances in machine learning (ML) have been facilitating numerous usecases
such as code generation~\cite{codex, copilot}, text generation~\cite{chatgpt, geminiteam2024gemini}, image generation~\cite{dalle3, diffusion, midjourney}, drug discovery~\cite{txllm}, just to name a few.
One of key trends in machine learning model development in the past few years is the sheer model size that has grown exponentially. When the attention mechanism~\cite{attention} was introduced in 2017, the original transformer model had 65 million parameters. Since then, the size of large language models (LLMs) has skyrocketed and trillion-parameter models (e.g., 1.6T Switch-C transformer~\cite{switchtrans}, 1.1T PanGu-$\Sigma$~\cite{ren2023pangusigma}, rumored Gemini~\cite{geminiteam2024gemini} and GPT-4~\cite{openai2024gpt4}) are not uncommon today.

These large models cannot fit into a single GPU or host, which requires distributed model deployment across GPUs and hosts to serve the models.
The distributed deployment necessitates a high-performance communication backend such as Nvidia's NCCL with NVLink and NVSwitch~\cite{nvlinkswitch} to maximize communication performance between GPUs or hosts. As a collective communication library (CCL), NCCL inherits a unique characteristic of CCL that makes it inadequate for serving the models efficiently while supporting fault tolerance and scalability.

More specifically, CCL cannot dynamically grow or shrink processes (workers) in a process group
(dubbed \emph{world}). Once a world is initialized, the departure of a worker breaks
the world. To recover the world, all active workers need to reinitialize a new 
world, which causes temporary unavailability of model serving.
The static nature of CCL also means that it cannot react to the sudden surge
of workloads quickly because adding workers in an existing world is impossible.
A workaround for this limitation is to create a new world and deploy the entire
model with more number of workers in the new world.
This can be expensive because a few partitions may be fully utilized while
others are not. In other words, partitions in a distributed deployment may have
different performances and thus there exist a few bottleneck partitions. In such
a case, a full model deployment can lead to the wastage of expensive GPU
resources.

We envision that a \emph{microservice}-like architecture at the partition level can
facilitate fault tolerance and scalability in a resource-efficient manner.
In this vision, partitions can scale out horizontally as the inference workloads
change over time without need for re-initializing all the workers.
This fine-grained scaling offers resource efficiency because overloaded partitions are only replicated.
As this architecture inherits the microservice's benefits, it can support fault tolerance naturally.
Therefore, overcoming the limitations of CCL is essential to realize the vision.

We design and develop \sys on top of PyTorch to overcome the limitations of CCL.
To best of our knowledge, \sys is the first framework that enables fault management and online scaling functionalities for CCL.
\sys allows one process to maintain multiple process groups, which offer the separation of fault domains.
One process (worker) failure only affects the process groups that it belongs to.
Other workers can tolerate the failure and continue to work within healthy process groups while cleaning up the broken process groups.
The support for multiple process groups also enables online scaling.
\sys allows new workers to join to an inference job by creating a new process group.
As inference workload varies over time, model serving systems can dynamically adjust their capacity by leveraging \sys's capability of supporting multiple process groups.
In this paper, we make the following contributions:

\begin{tightitemize}
  \item We propose \sys design (\secref{s:design}) that makes CCL elastic by leveraging asynchronous I/O and making fault detection and handling reliable. We develop fault tolerance and scaling functionalities on top of these mechanisms.
  \item We have open-sourced a \sys\footnote{\url{https://github.com/cisco-open/pymultiworld}} implementation (\secref{s:implement}) that is backward-compatible and easy to integrate.
    When PyTorch’s distributed collective operations are used, including a world name as a function argument suffices.
  \item Our preliminary evaluation (\secref{sec:eval}) shows that \sys is practical. It imposes only small overheads (1.4-4.3\% throughput loss) in most of the test scenarios.
\end{tightitemize}

\section{Motivation}
\label{s:motive}

Some key trends in ML drive the need for distributed model serving. First, the sheer model size has grown significantly over the past few years. Second, multiple-model composition is adopted to support complex workloads~\cite{modelcompose}. For instance, ensemble learning~\cite{ensemble_survey} involves multiple models to choose the best output.
These trends necessitate the inference job deployment across GPUs and hosts.

The model serving is an ``online'' and ``(soft)-realtime'' job and thus it is essential to make it elastic. We envision that elastic model serving requires two key functionalities: (i) online scaling, and (ii) fault management. The distributed ML deployment presents the characteristics similar to classical distributed systems. It should be able to offer horizontal scaling to serve time-varying inference requests resource-efficiently. Like the distributed systems equipped with fault management mechanisms to be robust against failures, the distributed model serving needs a proper fault management mechanism too.
Unsurprisingly, frequent failures (e.g., 1-2 GPU failures per week out of 400 GPUs, equivalent to a yearly failure rate of 13-26\%) are observed in ML clusters~\cite{bloomfailure, opt, megascale}.
Combined with other hardware failures, this frequent rate can cause serious SLO violations at the inference time.

The microservice architecture is well known for its ability that can satisfy both online scaling and fault management. In the architecture, service components scale out by launching more instances
in case of workload surge. Moreover, their deployment can grow and reduce independent of other components. This independent scaling allows the distributed systems to manage faults seamlessly without disruption.

Technologies such as message bus (e.g., Kafka) and queue (e.g., RabbitMQ) facilitate the microservice architecture. These technologies decouple service components by allowing communications between them. Applying the microservice architecture for the distributed model serving can naturally make it elastic. The stages (or partitions) of an inference job can scale out independently to absorb request spikes. The failures of the stages can be gracefully mitigated via redundancy and the performance can be restored via recovery mechanisms.

\begin{figure}[t]
\centering
\includegraphics[width=0.9\columnwidth]{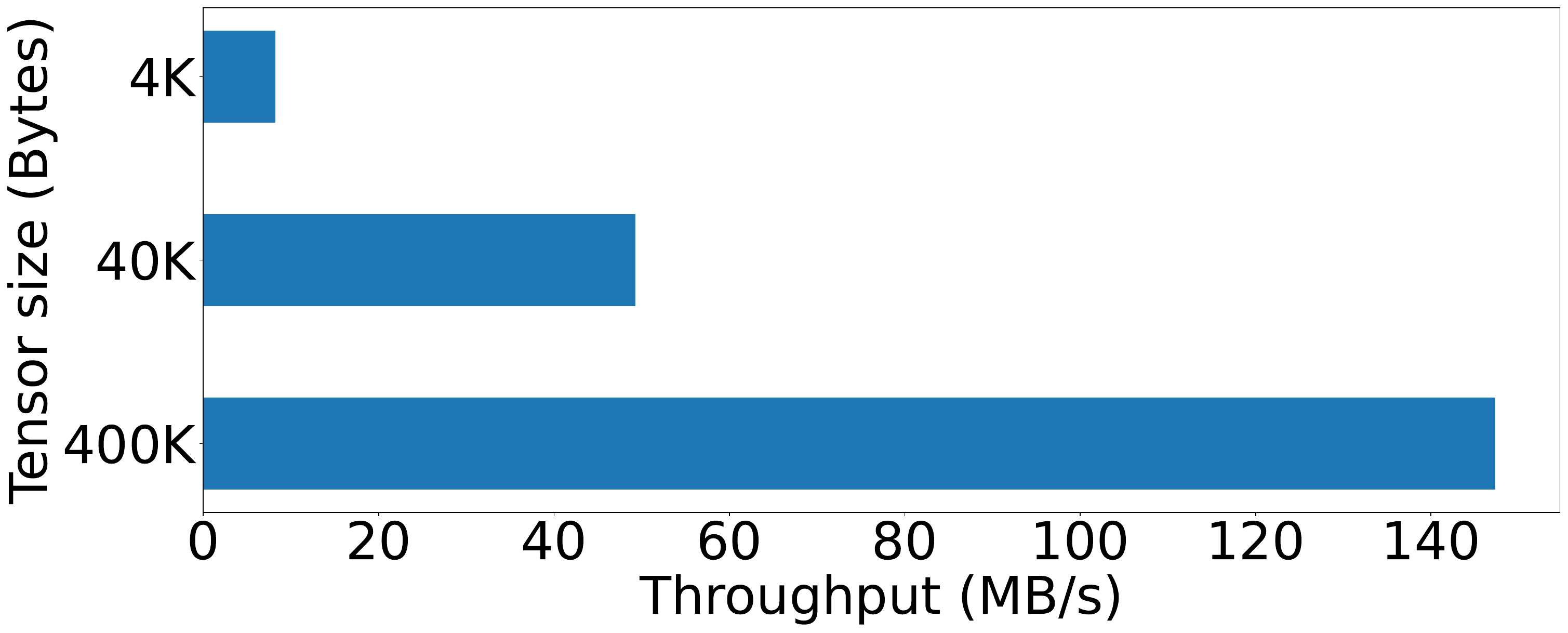}
\vspace{-.1in}
\caption{Throughput for tensor forwarding via Kafka.}
\label{fig:kafka}
\vspace{-.1in}
\end{figure}

Using message bus or queue for the distributed model serving, however, poses
high processing overheads. In the model serving, the information flows through a
neural network in the form of a tensor.
The message bus/queue technology makes it unavoidable to (de)serialize the tensors across
networks, and to copy them between CPU and GPU memory. These reduce throughput
greatly and slow down the serving speed. \fref{fig:kafka} illustrates the throughput for
sending tensors of various sizes across Kafka in our AWS testbed setup. We
observe that up to 45\% of the sender's time is spent copying the tensor from
GPU memory to CPU memory and then serializing it, while up to 53\% of the
receiver's time is spent reversing this process.
As shown in the figure, Kafka merely achieves about 147~MB/s when the tensor size is 400K. 

Collective communication libraries (CCLs) are crucial for high speed communication among ML training workers. Coupled with NVLink and NVSwitch~\cite{nvlinkswitch}, NCCL can achieve several hundreds of GB/s across GPUs and hosts. However, for model serving, CCLs fall short in supporting fault management and online scaling due to their static nature of creating a process group. Once a process group (dubbed \textit{world}) is created with a fixed number of processes (workers), the workers cannot leave the world. Thus, the failure of any worker leads to the restart of all active workers~\cite{nccldoc}, which is detrimental to high availability. The CCLs prevent adding new workers into an existing world. To adapt to fluctuating inference requests, it would be better to scale in and out individual, bottlenecked, stages. 
However, the fixed nature in the world size mandates the redundant deployment of an entire model, which can lead to more resource usage than stage-wise dynamic scaling to support the same amount of inference demands. Since GPUs are expensive resources, the resource efficiency of GPU-driven ML applications becomes more important than that of CPU-driven applications.

\begin{figure*}[t]
\centering
\subfloat[Normal state]{
    \includegraphics[width=0.31\textwidth]{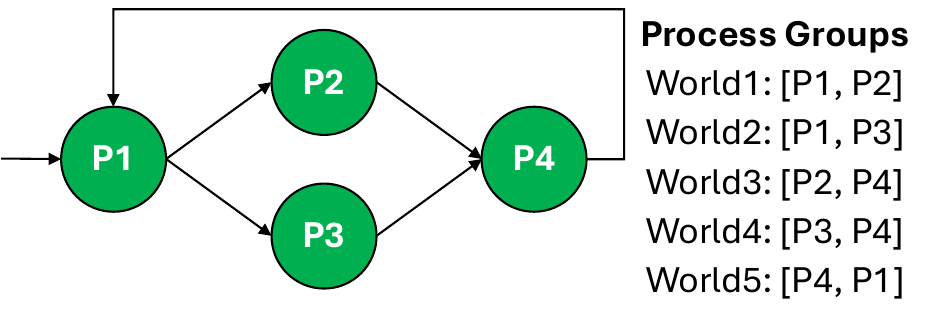}
    \label{fig:mw_ex1}
}
\subfloat[Worker failure]{
    \includegraphics[width=0.31\textwidth]{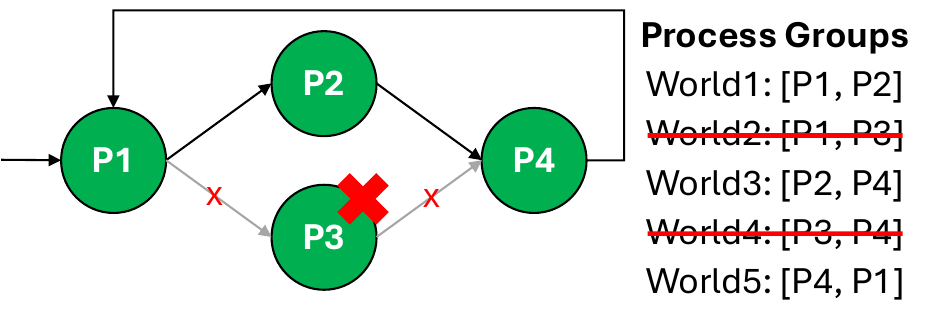}
    \label{fig:mw_ex2}
}
\subfloat[Failure recovery]{
    \includegraphics[width=0.31\textwidth]{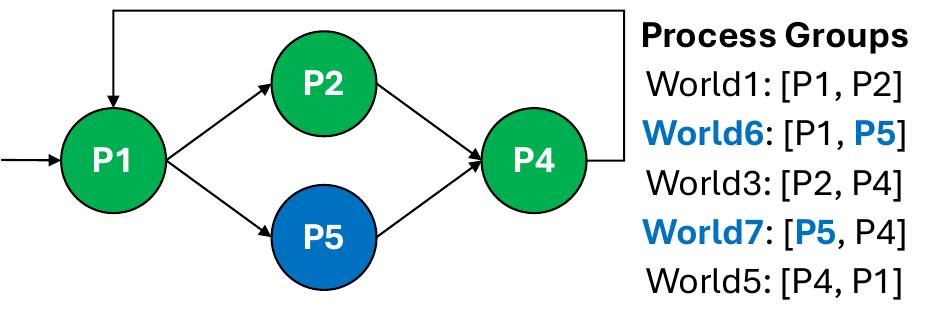}
    \label{fig:mw_ex3}
}
\vspace{-0.1in}
\caption{Illustration of \sys's elasticity. (a) A serving pipeline has three stages where the middle stage is replicated. (b) In case of a worker failure (here P3), worlds containing the failed worker are removed; and the remaining workers continue to work. (c) Online instantiation not only allows fault recovery without restarting all other workers, but it also enables online scaling.}
\label{fig:mw_ex}
\vspace{-0.1in}
\end{figure*}

\section{MultiWorld}
\label{s:multiworld}

\subsection{Overview}
\label{s:multiworld:overview}
\sys aims to enable primitives for fault management and online scaling
in a model serving system. The core idea of \sys is straightforward; to overcome
the limitation of the CCL's process group concept, \sys makes it possible for a
worker to join more than one world dynamically.
Workers in one world share a single fault domain, meaning that a worker
failure only affects the workers in the same world. Therefore, those affected
workers need to clean up their state and resources allocated for CCL. Allowing multiple
worlds for a worker makes fine-grained failure handling possible.
\sys can gracefully shut down the broken world(s) affected by a worker
failure. Since node failure can be translated into failures of workers running
in the node, \sys can handle node failure as well.

We illustrate how \sys can facilitate fault management and online scaling for model serving through \fref{fig:mw_ex}. Consider an ML model split into three partitions (or stages). One process is responsible for one stage. In the figure, we assume that the middle stage (stage 2) is a bottleneck and we replicate it with two processes--P2 and P3. As a result, the pipeline is constructed in a rhombus shape. We configure \sys to form a separate world for each edge between a pair of processes as depicted in \fref{fig:mw_ex1}. P2 and P3 can concurrently handle the outputs forwarded from P1. 

\mypara{Fault tolerance.} When a worker failure happens (e.g., P3 in \fref{fig:mw_ex2}), \sys helps the processes (P1 and P4) belonging to the same world of the failed worker detect the failure, and clean up the state and resources associated with the broken worlds (i.e., World2 and World4). P2 is not aware of P3's failure since P2 and P3 do not belong to any common world. P1 and P4 continue to work with P2. This is how \sys offers fault tolerance.

\mypara{Online instantiation.} Later, via a controller, a new worker (P5 in \fref{fig:mw_ex3}) can be created and added back to the existing pipeline by configuring P5 to inherit the exact role of P3 and other workers (P1 and P4) to set up new worlds with P5 (i.e., World6 and World7). We call this procedure online instantiation. This enables fault recovery as well as fine-grained online scaling (i.e., horizontal scaling at a stage level) for bottleneck stages. Note that the design and implementation of a controller is out of scope of this work, and we leave it as future work. Instead, our work focuses on functionalities to enable fault management and online scaling.

\subsection{Design}
\label{s:design}

We design \sys so that it can leverage different CCLs such as NCCL and GLOO. We focus on fault management and online scaling for NCCL due to its high performance. Using NCCL presents three key challenges. First, CCL operations should be non-blocking so that other tasks within a process can be scheduled; achieving this in a fault tolerance manner is non-trivial. Second, it is challenging to keep the low overhead of managing state for multiple worlds. A naive approach can lead to significant performance loss. Third, unreliable fault detection makes effective fault management hard. 

\mypara{Non-blocking CCL operations.} It is hard to make a serving system elastic with blocking CCL operations. For instance, P4 from \fref{fig:mw_ex1} should be able to receive outputs from P2 and P3 in an arbitrary order. If \texttt{recv} call for P2 blocks \texttt{recv} call for P3 or vice-versa, \sys can run into a deadlock. There are two ways to overcome this challenge: (i) threading blocking I/O, and (ii) asynchronous CCL operation.

The first option poses a practical drawback. Within a host, NCCL uses shared memory instead of networking stack. A blocking call does not get canceled in the presence of failure from other workers because no networking I/O event is raised.
Thus, handling the issue becomes complex.

On the other hand, the second option, asynchronous operation is free from the drawback of the first. It does not even require threading, which makes the design simple and less error-prone.
With asynchronous operation, however, the status of the operation is not informed, and thus a polling mechanism is required in order to check the status. If the status is not checked frequently, it can result in throughput loss. We mitigate the throughput loss of polling via busy waiting, but at the same time we make sure that other tasks can be scheduled immediately if the operation is pending. We discuss details on how to implement this feature in \secref{s:implement}.

\mypara{State management for multiple worlds.} It is essential to maintain separate state for different worlds so that communication can be enabled within any of the worlds a process belongs to. There are two ways to do this with PyTorch. One is to save the state information for each of the worlds outside of PyTorch and to restore the state of a world when communication needs to happen within the world. This is a time-based multiplexing approach that requires minimal changes on PyTorch. However, swapping the state information between worlds in a time-sharing manner costs \sys's performance, especially in case where the number of worlds increases. To eliminate the cost of the state swapping, we explore another approach in which each world's state is maintained as key-value pair within PyTorch. This approach is simple and effective, but requires additional modifications in PyTorch codebase. We plan to upstream those changes into PyTorch repository. Although our design is geared towards PyTorch, the lesson we learned here can be applicable even for other frameworks such as TensorFlow~\cite{tensorflow} and cupy~\cite{cupy}.

\mypara{Reliable fault detection.} NCCL uses two approaches for its CCL operations. It involves the OS networking stack for the host-to-host communication whereas it bypasses the networking stack and uses shared memory for GPU-to-GPU communication within a single host. When a failure happens under host-to-host communication, NCCL can catches the issue and throws \texttt{ncclRemoteError}. However, as mentioned earlier, the communication via shared memory does not raise any exception even in the presence of a failure, and thus the failure goes undetected. To detect fault effectively, we develop two approaches. Since NCCL throws \texttt{ncclRemoteError}, \sys catches and handles the exception gracefully. For the shared-memory case, we develop watchdog to monitor workers' liveness. We discuss more details on watchdog in \secref{s:implement}.

\begin{figure}[t]
\centering
\includegraphics[width=0.85\columnwidth]{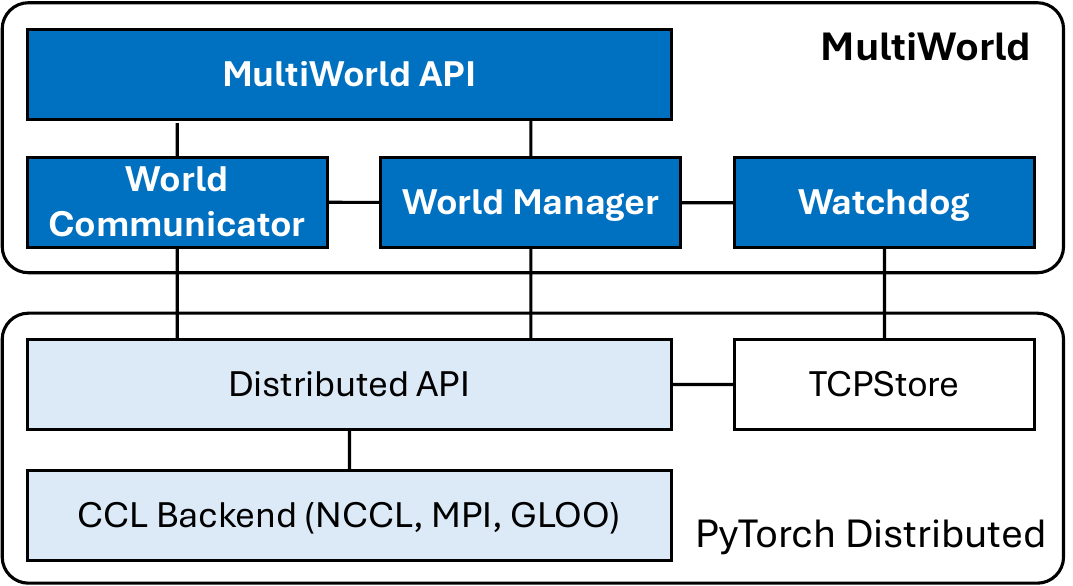}
\vspace{-.1in}
\caption{\sys Architecture.}
\label{fig:arch}
\end{figure}

\begin{figure}[t]
\centering
\subfloat[Single world]{
    \includegraphics[width=0.42\columnwidth]{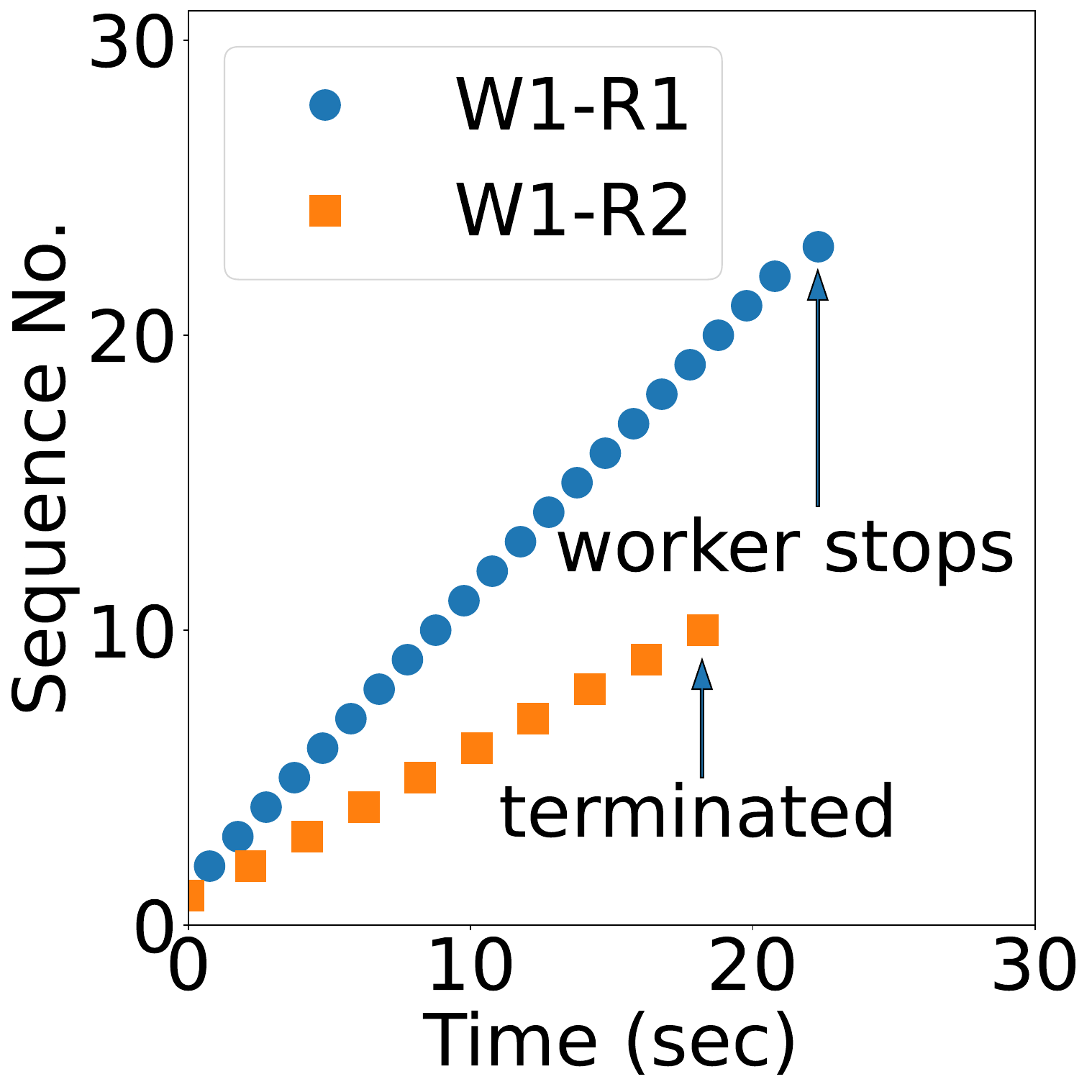}
    \label{fig:fault_intol_ex}
}
\hfill
\subfloat[\sys]{
    \includegraphics[width=0.42\columnwidth]{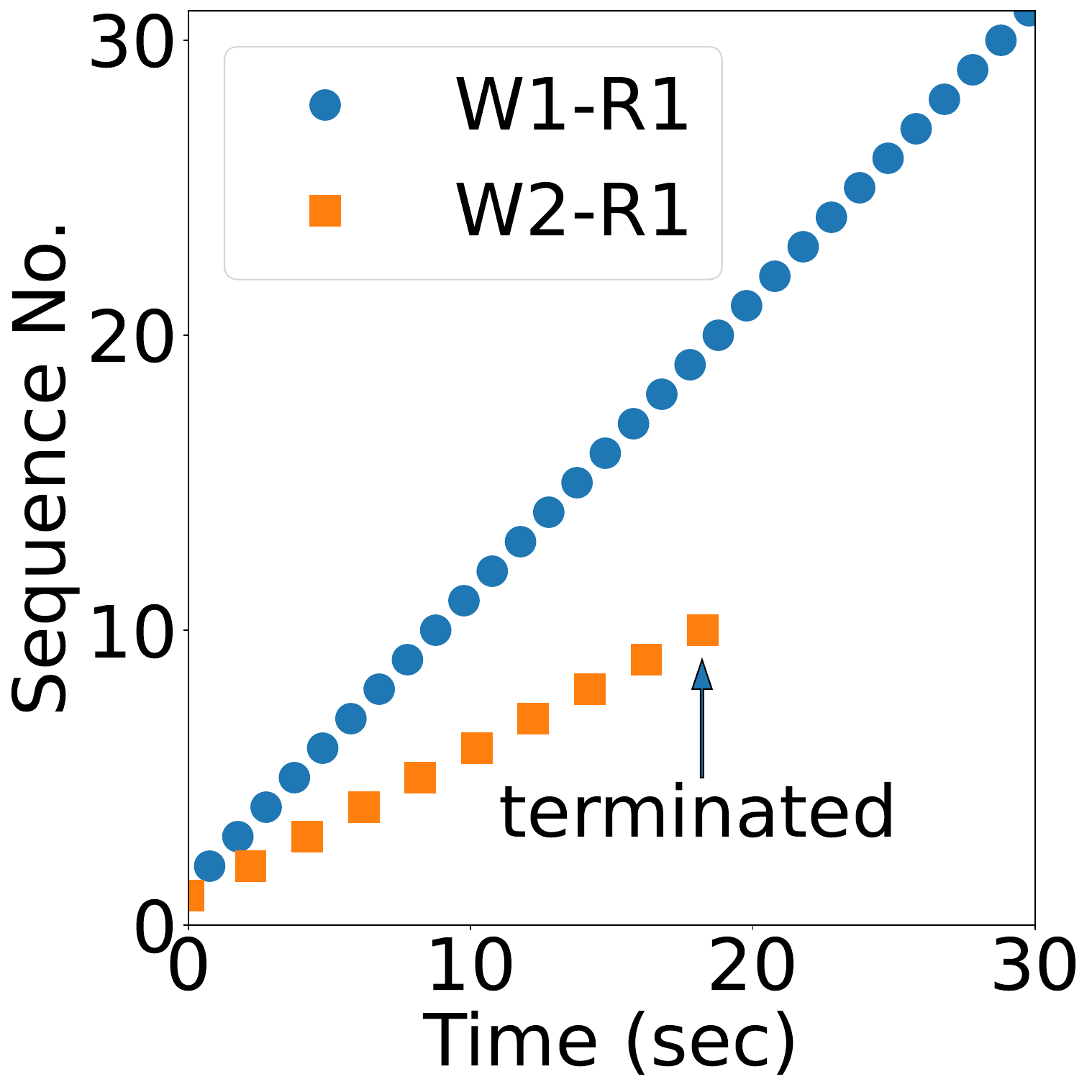}
    \label{fig:fault_tol_ex}
}
\vspace{-.1in}
\caption{Fault tolerance of \sys.
Across two cases (single world and \sys), a worker gets terminated after sending the 10th tensor. In the single world case (left), the other worker stops working once it detects the failure. In case of \sys (right), the other worker continues to operate successfully.}
\label{fig:fault_example}
\vspace{-.1in}
\end{figure}

\subsection{Implementation}
\label{s:implement}

We implement \sys with about 600 lines of code on top of PyTorch v2.2.1. We also modify PyTorch's distributed module to support multiple worlds.
\sys mainly consists of three components: (i) world manager, (ii) world communicator, and (iii) watchdog. \fref{fig:arch} depicts how \sys is architected with these three components.

\mypara{World Manager.} This component manages initialization and termination of a world. If the watchdog alerts a world's failure, the manager prevents the broken world being accessed by the world communicator. It then helps the communicator abort any pending collective operation and raise an exception so that an inference application can handle it. It provides three functions: \texttt{initialize\_world}, \texttt{remove\_world}, and \texttt{communicator}. Since the first two functions are self-explanatory, we only explain \texttt{communicator} briefly; it is a function
that returns an object of the world communicator. With this function, the serving application can use the collective operations offered by the communicator.

\mypara{World Communicator.} This component offers a set of fault-tolerant collective operations. It allows inference applications to issue the collective operations across different worlds in a non-blocking fashion by leveraging asynchronous mode in PyTorch distributed module and Python's \texttt{asyncio} module. This addresses the potential deadlock issue discussed in \secref{s:design}.
The busy waiting method discussed in \secref{s:design} requires 100\% utilization of only one CPU core. We trade one CPU core for fault tolerance and online scaling. The dedication of one CPU core is relatively cheap since we can increase GPU utilization by putting workloads onto GPU as fast as possible. We support 8 collective operations: \texttt{send}, \texttt{recv}, \texttt{broadcast}, \texttt{all-reduce}, \texttt{reduce}, \texttt{all-gather}, \texttt{gather}, and \texttt{scatter}.

\mypara{Watchdog.} It is a threaded daemon that checks whether worlds that a worker belongs to are broken or not. It relies on TCPStore created by PyTorch during the initialization of a world. One TCPStore instance is associated with one world. A watchdog updates the worker's health periodically to the stores for all the worlds the worker belongs to. If health updates are missed for a certain duration (e.g., 3~seconds), the watchdog informs the world manager, which in turn cleans up the state and resources for the broken world.

\section{Evaluation}
\label{sec:eval}

We evaluate \sys in terms of fault tolerance, online instantiation and performance. We use two p3.8xlarge VMs, each of which has four Nvidia's Tesla V100 GPUs interconnected with NVLink.
The bandwidth between the VMs is 10~Gbps. We use the following notation across all experiments to specify a process identifier, W$x$-R$y$ where $x$ denotes a process group (world) of the process and $y$ is its rank within the world. In case of \sys, a process can belong to more than one world, and thus it can have more than one identifier. We use NCCL
for our evaluation.

\begin{figure}[t]
\includegraphics[width=0.9\columnwidth]{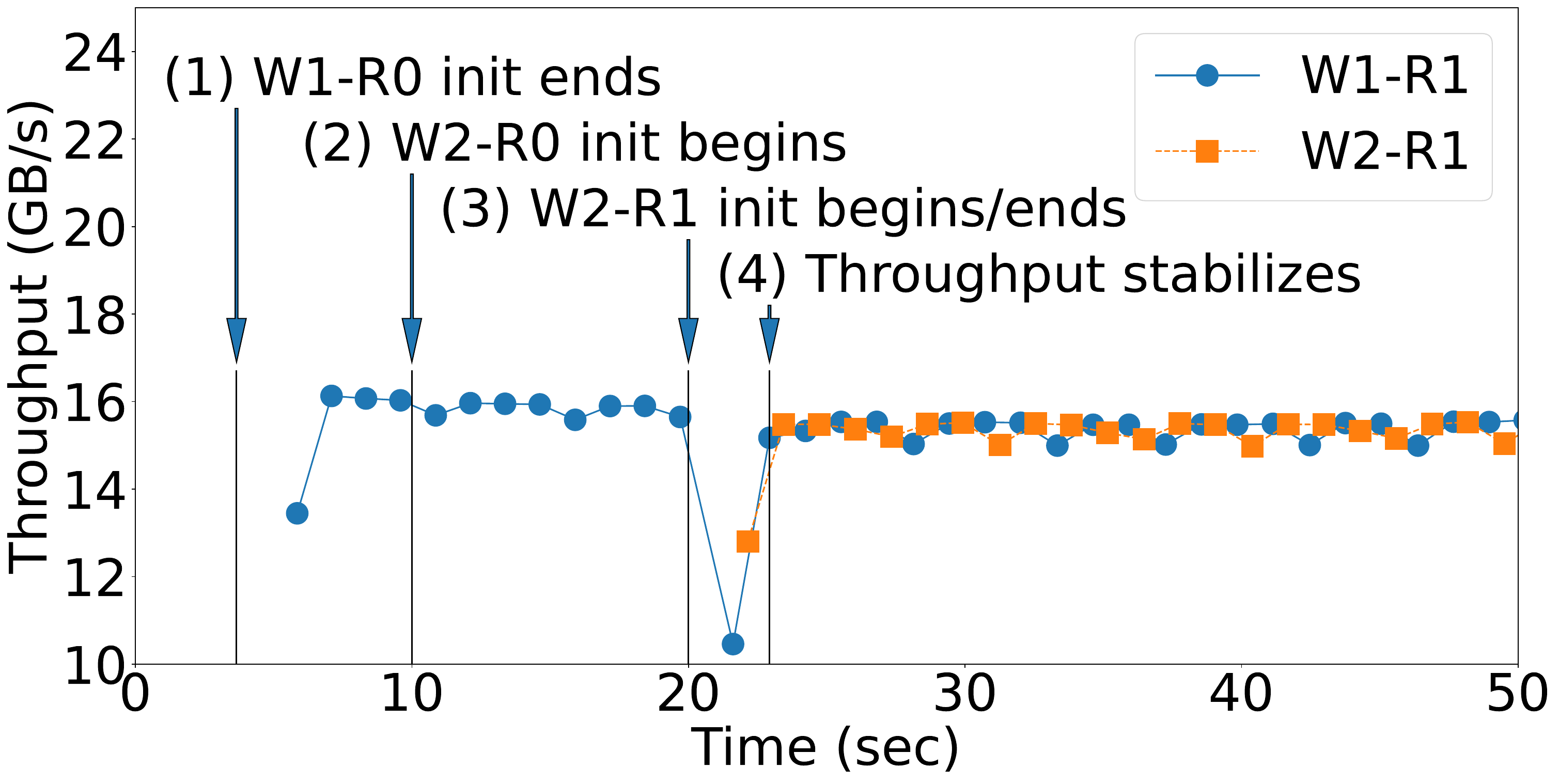}
\vspace{-.1in}
\caption{Adding a worker dynamically.}
\label{fig:mw_scaling_thp}
\vspace{-.1in}
\end{figure}

\subsection{Fault Tolerance}
\label{s:fault}

We launch three processes across two machines. One process is a leader process and the other two are workers. For \sys, we assign W1-R0 and W2-R0 to the leader process as its identifiers. The other two have W1-R1 and W2-R1. Note that a process can be a leader for one world but a worker for another. For a single world, the three processes have W1-R0, W1-R1, and W1-R2 as their identifier. We deploy the leader process in one host and the other two in the other host. W1-R1 sends the leader process one message per second while W1-R2 and W2-R1 send the message every two seconds. We terminate W1-R2 and W2-R1 after they send the 10th message. \fref{fig:fault_example} shows how the leader process in \sys and the single world handles a worker failure. In \fref{fig:fault_intol_ex}, the leader continues to receive a couple of more tensors from W1-R1, but eventually stops receiving tensors (at 22.3~sec mark). On the other hand, in \fref{fig:fault_tol_ex}, the leader and W1-R1 continue to work successfully, which demonstrates \sys's fault tolerance capability.

\begin{figure}[t]
\centering
\subfloat[GPU to GPU]{
    \includegraphics[width=0.47\columnwidth]{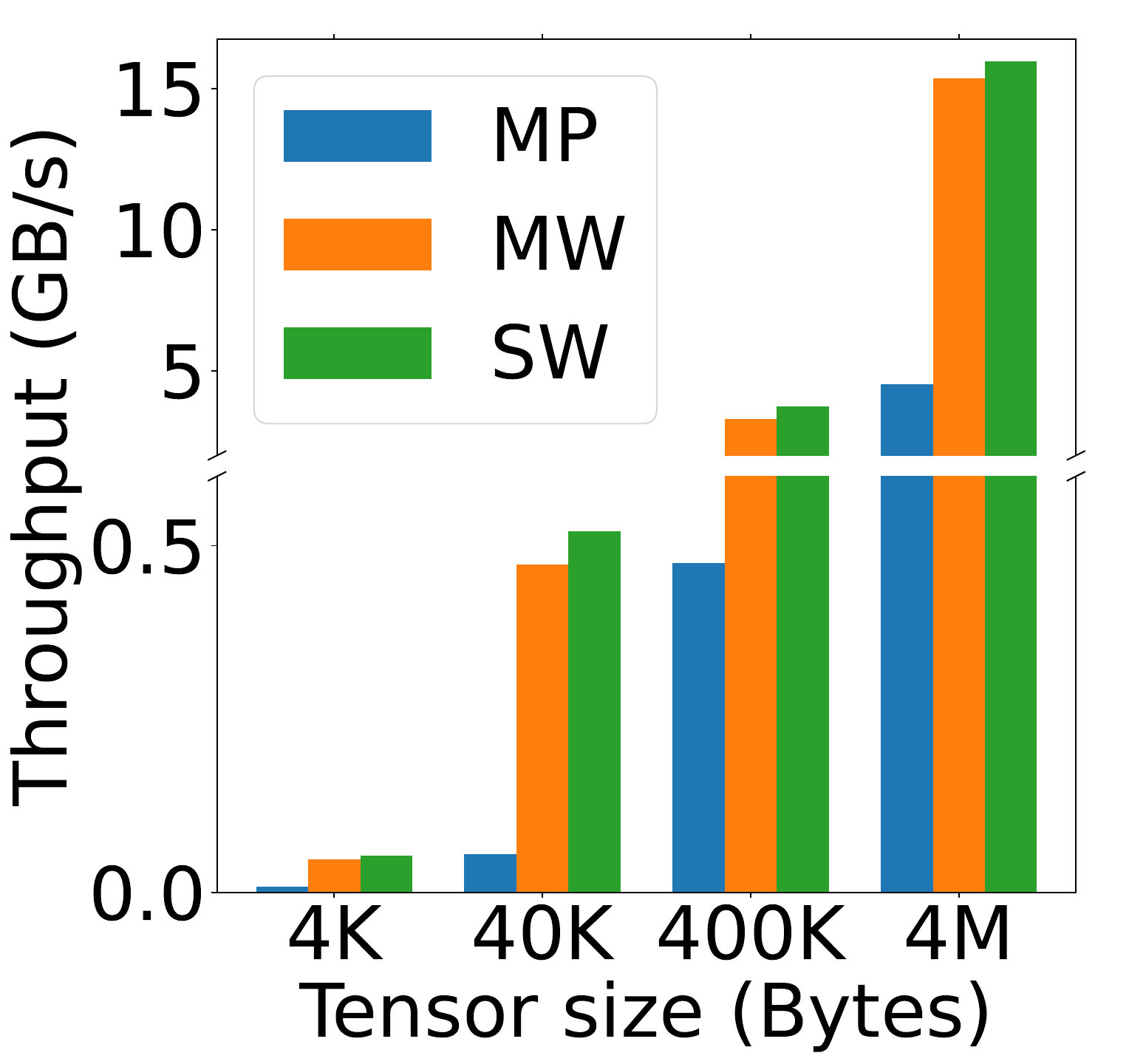}
    \label{fig:nccl_thp_sh}
}
\hfill
\subfloat[Host to host]{
    \includegraphics[width=0.43\columnwidth]{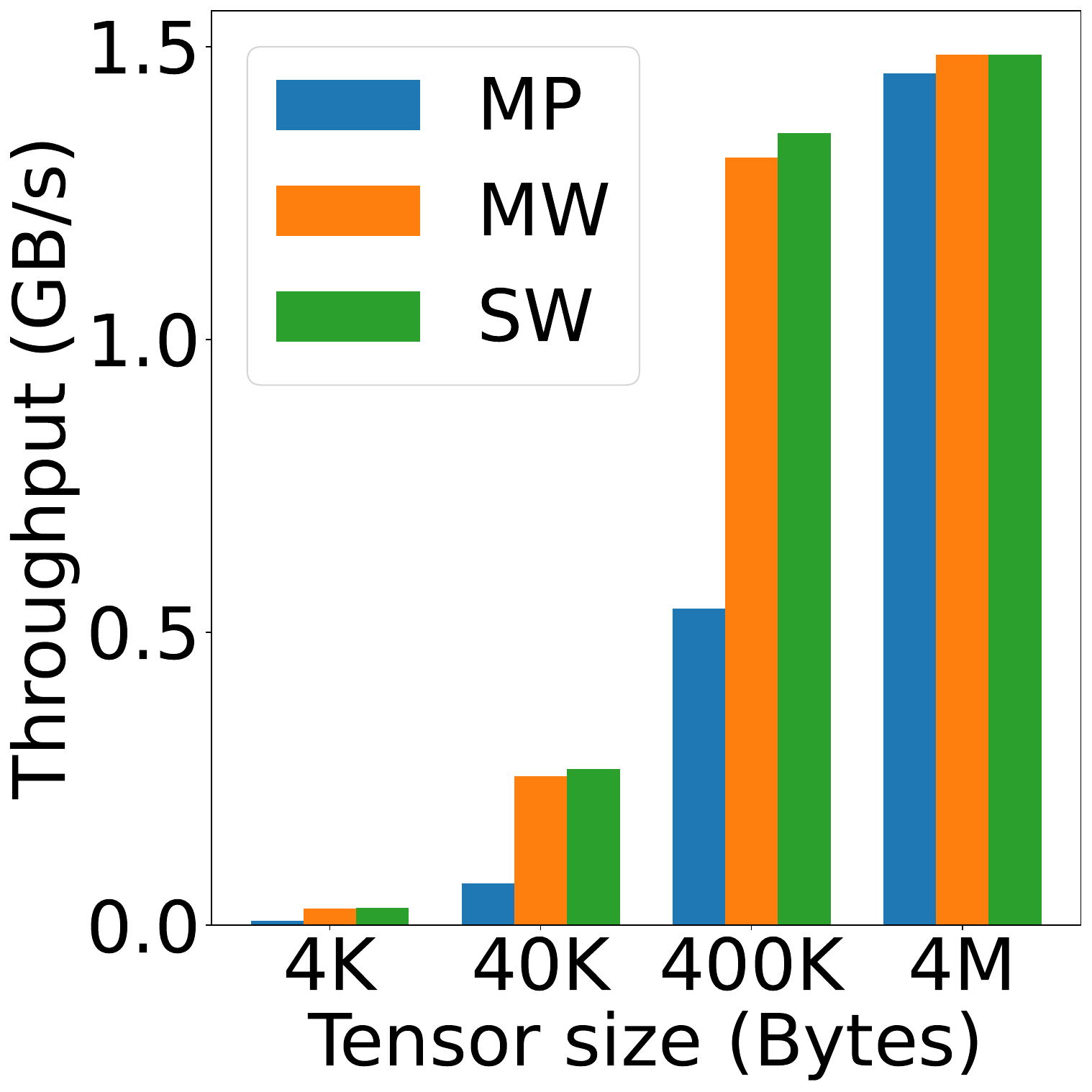}
    \label{fig:nccl_thp_twohost}
}
\vspace{-.1in}
\caption{Throughput comparison. MP, MW and SW denote MultiProcessing, \sys and single world, respectively. SW is built on top of vanilla PyTorch.}
\label{fig:single_send_recv_thp}
\vspace{-.1in}
\end{figure}

\begin{figure*}[t]
\centering
\subfloat[Tensor size = 4KB]{
    \includegraphics[width=0.48\columnwidth]{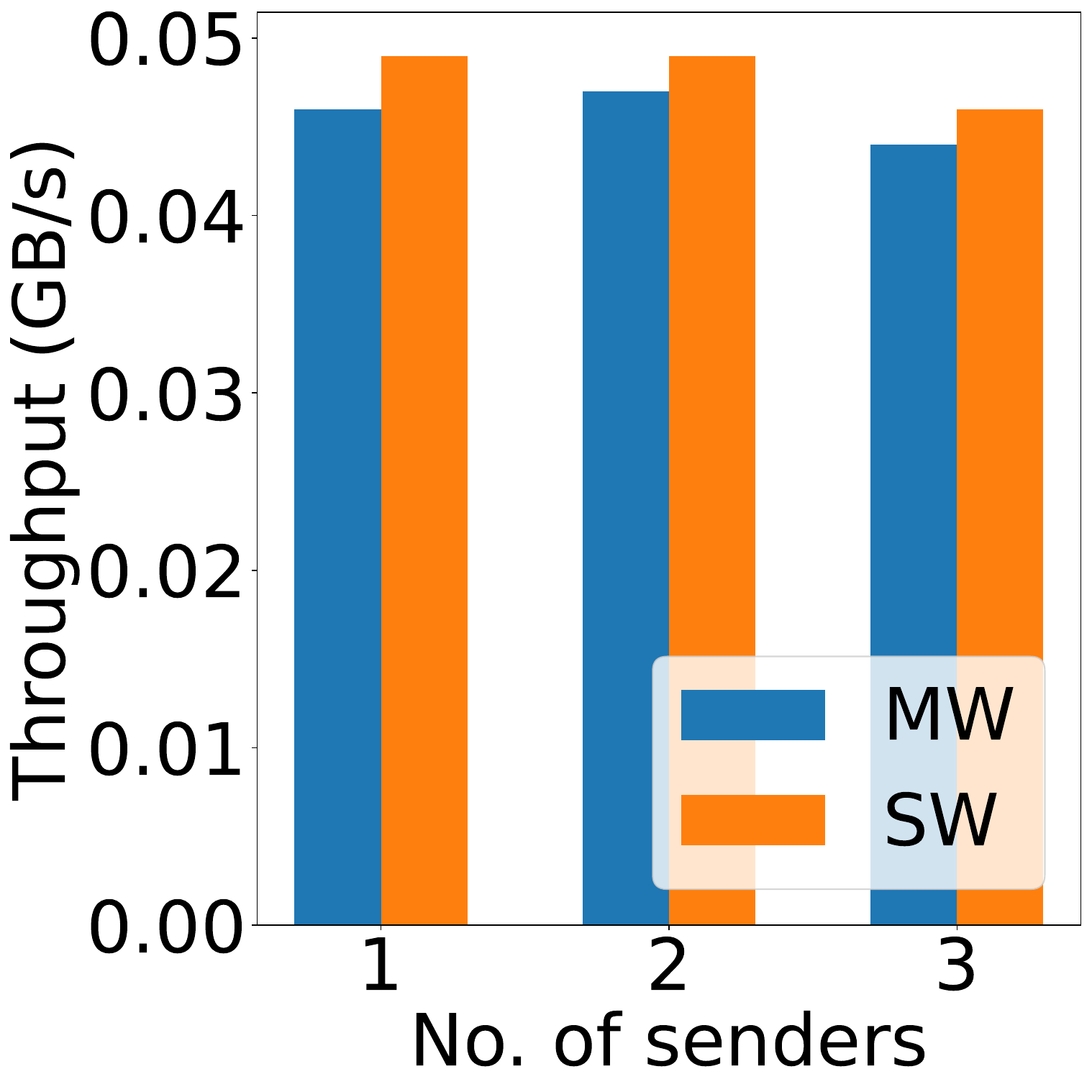}
    \label{fig:nccl-thp-sh-multisender-tsize1000}
}
\hfill
\subfloat[Tensor size = 40KB]{
    \includegraphics[width=0.48\columnwidth]{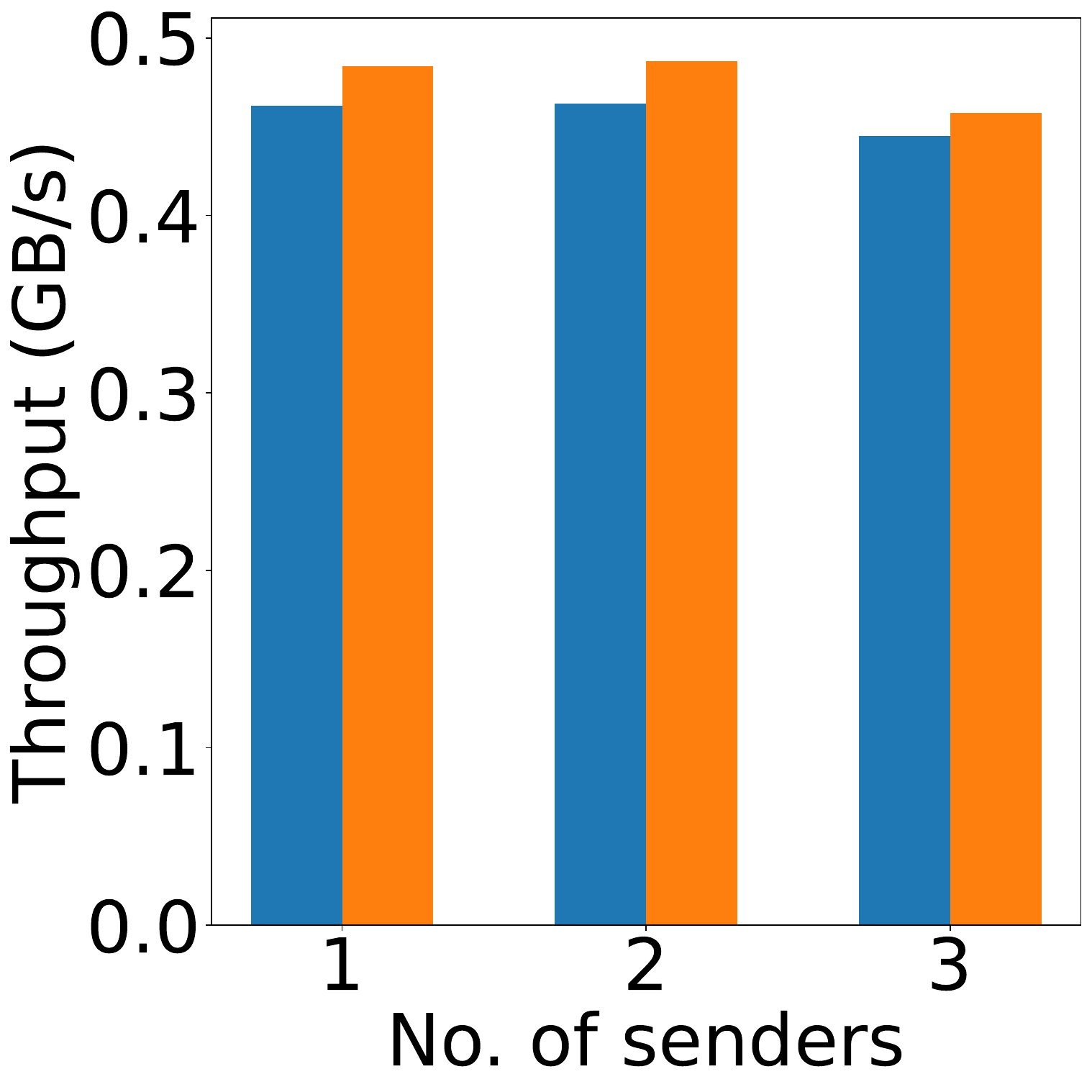}
    \label{fig:nccl-thp-sh-multisender-tsize10000}
}
\hfill
\subfloat[Tensor size = 400KB]{
    \includegraphics[width=0.48\columnwidth]{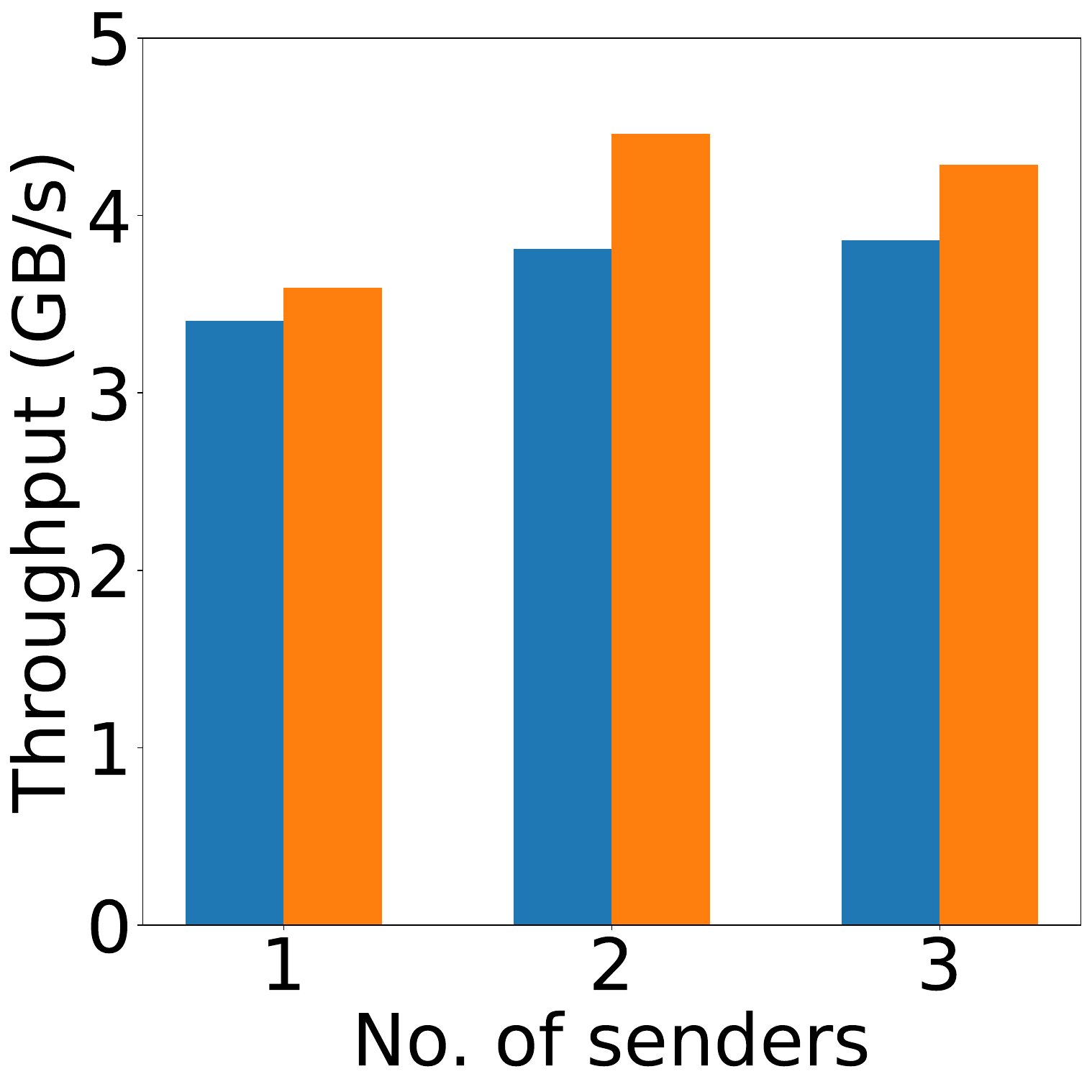}
    \label{fig:nccl-thp-sh-multisender-tsize100000}
}
\hfill
\subfloat[Tensor size = 4MB]{
    \includegraphics[width=0.48\columnwidth]{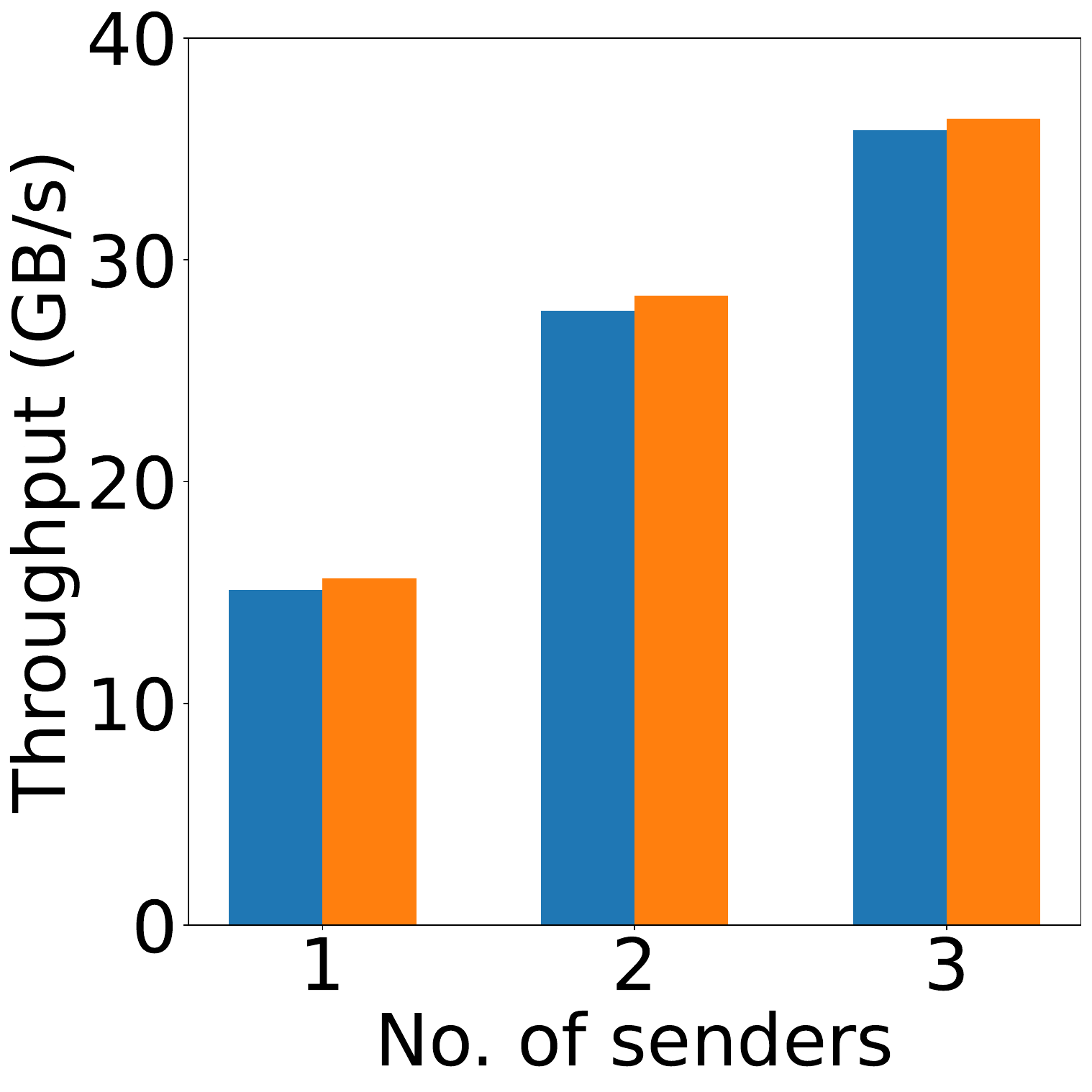}
    \label{fig:nccl-thp-sh-multisender-tsize1000000}
}
\vspace{-.1in}
\caption{GPU-to-GPU throughput of one receiver by varying the number of senders.}
\label{fig:multisender_thp}
\vspace{-.1in}
\end{figure*}

\subsection{Online Instantiation}

Another crucial functionality is an ability to scale out workers without restarting all the workers in a world when inference demands increase. To demonstrate the online instantiation ability of \sys, we use the setting in \secref{s:fault} with a few modifications. We deploy the three processes in a single host to run the experiment with NVLink.
The tensor size is set to 4~MB (a 32-bit floating point tensor whose length is 1M). \fref{fig:mw_scaling_thp} depicts how the online instantiation impacts the throughput of the existing communication. The leader finishes the initialization of W1 at 3.66~sec mark and starts to receive the tensors. We compute the throughput of the leader process from its respective worker in a different world every time it receives 5,000 tensors.

At the beginning (at 5.8~sec mark), the leader achieves 13.4~GB/s and the throughput stabilizes at 15.9~GB/s in the subsequent data points. The initial lower throughput seems to stem from the fact that PyTorch initializes NCCL's communicator in a lazy fashion (i.e., the communicator is created upon the first collective operation such as \texttt{isend} or \texttt{irecv}), and the communicator creation is a collective operation. We let the leader initialize W2 at 10~sec mark while W2-R1 joins W2 at 20~sec mark. The joining step only takes 20~ms. There is no impact on W1-R1's throughput between the 10~sec and 20~sec marks even if the leader is waiting for W2-R1's arrival. This is because \sys handles this blocking initialization in a separate thread in a thread-safe manner. As W2-R1 starts to send the tensors to the leader (at 20~sec mark), the throughput for both W1-R1 and W2-R1 is lower than 15.9~GB/s. Again, this seems to be due to the lazy initialization of PyTorch. After the initial degradation period (at 22.9~sec mark), both workers achieve about 15.4~GB/s. This evaluation shows that \sys can dynamically scale out workers with small performance penalty.

\subsection{Throughput}

We measure the throughput performance across three different scenarios to demonstrate that the overhead of \sys is small compared to the performance of the vanilla PyTorch distributed package. We run the experiments 10 times and present the average throughput.

\mypara{GPU-to-GPU.} We first present the GPU-to-GPU communication overhead of \sys under the scenario of one sender and one receiver. The baseline is the performance of the vanilla PyTorch distributed package. We also implement an alternative architecture of \sys, in which a main process leverages multiple sub-processes, each of which manages a different world. We call this architecture MultiProcessing (MP). In MP, tensors are moved via PyTorch's IPC mechanism (e.g., pipe or queue in \texttt{torch.distributed}) between the main process and a sub-process. \fref{fig:nccl_thp_sh} depicts that \sys's throughput is close to that of the single world (SW) implementation. Hence, \sys's overhead is small. On the other hand, MP performs poorly, especially when tensor size is small ($\le 400K$). Even when the tensor size is 4~MB, MP only achieves 4.53~GB/s, which is only 30\% of the throughput achieved by \sys or SW. This is mainly because of the overhead of the IPC.

\mypara{Host-to-host.} \fref{fig:nccl_thp_twohost} shows the similar trend observed in the GPU-to-GPU communication setting. While MP still suffers in case of the small tensor size, its performance becomes comparable to other two methods, \sys and SW, in case of the 4~MB tensor. In contrast, \sys works as well as SW across different tensor sizes. They can saturate the 10~Gbps bandwidth as the tensor size increases.

\mypara{Multiple senders and one receiver.} We now measure the aggregate throughput by varying the number of senders from one to three since our VM has 4 GPUs.
We also vary the tensor sizes from 4~KB to 4~MB.
\fref{fig:multisender_thp} shows that \sys only incurs small overheads (1.4-4.3\%) in the most of the cases. We observe from \fref{fig:nccl-thp-sh-multisender-tsize100000} that \sys faces 14.6\% (3.81~GB/s vs 4.46~GB/s) throughput degradation in the worst case. However, as the tensor size increases, the overall overhead becomes negligible (\fref{fig:nccl-thp-sh-multisender-tsize1000000}).

\section{Related Work}
\label{sec:related}

\paragraph{Model serving systems:}
Existing systems have various approaches to fault tolerance and failure
resilience. Orca~\cite{orca} does not address fault tolerance in either its implementation
or design.
Similarly, Clipper~\cite{clipper} focuses on model selection rather than fault tolerance, improving
failure resilience through model redundancy but not addressing execution
resilience for single query inference. AlpaServe~\cite{alpaserve}, built on the Alpa~\cite{alpa},
lacks explicit failure recovery mechanisms.
AlpaServe's model splitting across workers and reliance on static communication
groups introduce failure-prone challenges that our work aims to mitigate.
STI~\cite{sti}, designed for low-latency and resource-constrained environments
like mobile devices, adjusts model size to meet latency requirements. 
SpotServe~\cite{spotserve} is designed to leverage low-cost spot GPU instances
for LLM inference to maintain service continuity through context and progress
migration, and \sys-like design would facilitate its implementation.

\paragraph{Scalable training systems:}
Elastic Horovod~\cite{elastichorovod}, FfDL~\cite{ffdl} and several other frameworks are
designed
for scalable training.
Horovod Elastic allows for dynamic
resizing of the training cluster, enabling the addition or removal of workers
during training without restarting the whole job. This elasticity enhances
resource utilization and fault tolerance, as training can continue even if some
nodes fail or become unavailable. 
Similar is the functionality of FfDL.
Developer's have introduced experimental feature of dynamic world size in
RPC~\cite{drpc} in PyTorch to allow training systems to adjust the number of
participating
nodes dynamically, facilitating more resilient and efficient training processes.
Oobleck~\cite{oobleck} ensures fault-tolerant distributed training of large DNN
models by using heterogeneous pipeline templates and replicated model states for
fast recovery.

\section{Conclusion}

In this paper we design and develop \sys that enables fine-grained fault tolerance and online scaling to make collective communication library (CCL) more suitable for model serving.
At the core of \sys are three ideas: (i) asynchronous CCL operations, (ii) efficient multi-process group management, and (iii) reliable failure detection.
We demonstrate that \sys supports the functionalities with small overheads through various real experiments.
We believe that \sys is a solid building block to facilitate the development of a full-fledged model serving system that can be resilient against failures and scale inference jobs elastically in response to dynamic demand changes. We plan to build such a system as future work.

\section*{Acknowledgments}
We would like to thank Pranav Gadikar for contributing to the implementation of the framework at the early stage of the project.

\balance
\bibliographystyle{ACM-Reference-Format} 
\bibliography{main}

\end{document}